\documentstyle[prd,aps,floats]{revtex}
\input{epsf}
\def\to{\rightarrow}
\def\gev{\mbox{GeV}}
\def\ev{\mbox{eV}}
\def\mev{\mbox{MeV}}

\def\mpc{\mbox{Mpc}}
 
\def\AJ{{\it Ap. J.} }
\def\AJL{{\it Ap. J. Lett.} }

\def\APJ{{\it Ap. J.} }

\def\EURO{{\it The European Phys. J.} }

\def\MNRAS{{\it Mon. Not. R. Ast. Soc.} }
\def\NAT{{\it Nature} }

\def\NP{{\it Nucl. Phys.} }
\def\PL{{\it Phys. Lett.} }
\def\PR{{\it Phys. Rev.} }
\def\PRL{{\it Phys. Rev. Lett.} }

\def\SC{{\it Science} }

\def\vev#1{\langle {#1}\rangle}

\def\frac#1#2{{\textstyle{{#1}\over {#2}}}}

\def\lsim{\mathrel{\rlap{\lower4pt\hbox{\hskip1pt$\sim$}}
    \raise1pt\hbox{$<$}}}
\def\gsim{\mathrel{\rlap{\lower4pt\hbox{\hskip1pt$\sim$}}
    \raise1pt\hbox{$>$}}}
\def\sqr#1#2{{\vcenter{\vbox{\hrule height.#2pt
         \hbox{\vrule width.#2pt height#1pt \kern#1pt
         \vrule width.#2pt}
         \hrule height.#2pt}}}}

\newcommand{\beq}{\begin{equation}}
\newcommand{\eeq}{\end{equation}}
\newcommand{\bea}{\begin{eqnarray}}
\newcommand{\eea}{\end{eqnarray}}

\begin{document}
\draft
\twocolumn[\hsize\textwidth\columnwidth\hsize\csname 
@twocolumnfalse\endcsname
\preprint{DF/IST-7.2000\\
October 2000\\}

\title{Cosmological Constraints on an Invisibly Decaying Higgs}

\author{M.C. Bento $^{(1)}$, O. Bertolami $^{(1)}$ and  
R. Rosenfeld $^{(2)}$}

\address{(1) Instituto Superior T\'ecnico\\ Departamento de F\'\i sica\\
Av. Rovisco Pais 1, 1049-001 Lisboa, Portugal\\}

\address{(2) Instituto de F\'\i sica Te\'orica\\
R.\ Pamplona 145, 01405-900 S\~ao Paulo - SP, Brazil\\}

\address{E-mail addresses:  bento@sirius.ist.utl.pt; orfeu@cosmos.ist.utl.pt; 
rosenfel@ift.unesp.br}

\vskip 0.5cm

\date{\today}

\maketitle

\begin{abstract}
Working in the context of a proposal for collisional dark matter, we derive 
bounds on the Higgs boson coupling $g^{\prime}$ to a stable light scalar
particle, which we refer to as phion ($\phi$), required to solve problems
with small scale structure formation which arise in collisionless dark matter
models. We discuss the behaviour of the phion in the early universe for
different ranges of its mass. We find that a phion in the mass range of 
$100 ~\mev$ is excluded and that a phion in the mass range of $1 ~\gev$
requires a large coupling constant, $g^{\prime} \gsim 2$, and
$m_h \lsim 130~\gev$ in order to avoid overabundance, in which case the
invisible decay mode of the Higgs boson would be dominant. 

\vskip 0.5cm
 
\end{abstract}


\vskip 2pc ]
\section{Introduction}
\label{sec:int}

We have recently discussed the role of a light, stable, strongly self-coupled 
scalar field in solving the problems of the Cold Dark Matter (CDM) model  
for structure formation in the Universe, concerning galactic 
scales \cite{Bento1}. 
Our proposal involves a particle physics-motivated model, 
where the DM particles are allowed to self-interact so as to have a large 
scattering 
cross section and negligible annihilation or dissipation. The self-interaction
results in a characteristic length scale given by the mean free 
path of the particle in the halo.
This idea was originally proposed to suppress small scale power 
in the standard CDM model  
\cite{Carlson,Laix} and has been recently revived, in a general context, 
in order to address the issues discussed above \cite{Spergel}. 
Our model \cite{Bento1}  is 
a concrete realization of this idea, which involves an extra gauge singlet 
as the self-interacting, non-dissipative cold 
dark matter particle. Following Ref. \cite{Binoth}, we call this scalar 
particle phion, $\phi$, and assume that it couples
to the Standard Model (SM) Higgs boson, $h$, with a Lagrangian 
density given by:

\begin{equation}
{\cal L} = {1 \over 2} (\partial_\mu \phi)^2 - {1 \over 2} m_\phi^2 \phi^2  
- {g \over 4!} \phi^4 + g^{\prime} v \phi^2 h 
\quad,
\label{1}
\end{equation}%
where $g$ is the phion
self-coupling constant, $m_\phi$ its mass, $v=246\ \gev$ is the Higgs 
vacuum expectation value and $g'$ is the coupling between
$\phi$ and $h$. A model along these lines 
have been previously discussed \cite{Silveira}. Clearly the interaction term 
between the phion and the Higgs boson arises from a quartic interaction 
${g^{\prime} \over 2}  \phi^2 H^2$, where $H$ is the electroweak Higgs doublet.
As shown in  \cite{Bento1}, the $\phi$ mass does not arise 
from spontaneous symmetry breaking since this would yield a tiny scalar 
self-coupling constant. The phion mass in (\ref{1}) should be regarded as 
as a phenomenological parameter arising from a more encompassing theory.
    
As is well known, scalar particles 
have been repeatedly invoked as DM
candidates \cite{Gradwohl,McDonald,Bertolami,Peebles,Goodman,Matos}; 
however, our 
proposal has the salient feature that it brings about a 
connection with the SM Higgs boson which could arise in extensions of the
SM. For instance, the hidden sector of heterotic string theories does 
give rise to 
astrophysically interesting self-interacting scalars \cite{Faraggi}. 
For reasonable values of $g^{\prime}$, the new scalar
would introduce a novel invisible
decay mode for the Higgs boson. This could, in principle, provide an
explanation for the failure in finding the Higgs boson at accelerators
sofar\cite{Bij}, which will be tested at future colliders.

On the astrophysical front, recent observational data on large scale 
structure, cosmic microwave background 
anisotropies and type Ia supernovae suggest that $\Omega_{tot} \approx 1$, 
of which $\Omega_{baryons} \approx 0.05$ and $\Omega_\Lambda \approx 0.65$ 
\cite{Bahcall}; the remaining contribution, $\Omega_{DM}\approx 0.3$ 
(apart from neutrinos that may contribute a small fraction), comes from 
dark matter (DM), which
determines the hierarchy of the structure formation in the Universe. 
The most prominent theories of structure formation are now $\Lambda$CDM 
and QCDM, which consist, respectively, of the standard 
Cold Dark Matter (CDM) model 
supplemented by a cosmological constant or a dark energy, i.e. a negative 
pressure component.

In the  CDM model, initial
Gaussian density fluctuations, mostly in non-relativistic collisionless
particles, the so-called cold dark matter, grow during the  inflationary 
period of the Universe and evolve, via  gravitational instability, into the 
structures one observes at present.
However, it has been found that the CDM model cannot sucessfully accomodate 
the data observed  on all scales. For instance, N-body simulations 
predict a number of halos 
which is a factor $\sim$ 10 larger than the 
observed number at the level of Local Group \cite{Mooreetal2,Klypin}.
Furthermore, astrophysical systems that are DM dominated, e.g.  
dwarf galaxies \cite{Moore,Flores-Primack,Burkert1} and low surface brightness 
galaxies \cite{Blok} show shallow matter--density profiles with 
finite central densities. This is in contradiction with galactic 
and galaxy cluster halos in high resolution N-body simulations 
\cite{Navarro,Ghigna0,Ghigna,Mooreetal}, which have singular cores, 
with $\rho \sim r^{-\gamma}$ and $\gamma$ in the range between 1 and 2.
This can be understood as cold collisionless DM particles do not have 
any characteristic length scale leading, due to hierarchical gravitational 
collapse, to dense dark matter halos with negligible 
core radius.

It is relevant to stress that recent numerical simulations   
\cite{Hannestad,Moore1,Yoshida,Wandelt} indicate that self-interaction
of DM particles does bring noticeable 
improvements on properties of the CDM model on small scales. 
We point out, however, that a numerical simulation that takes into 
account the salient features of our proposal is still missing.

At present, $\phi$ particles are non-relativistic, with typical 
velocities 
$v \simeq 200 $ km s$^{-1}$, and, therefore, it is impossible to dissipate
energy creating more particles in reactions like 
$\phi \phi \rightarrow \phi \phi \phi
\phi$. Thus, as only the elastic channel is kinematically accessible, near 
threshold, the cross section is given by:

\begin{equation}
\sigma (\phi \phi \rightarrow \phi \phi) \equiv \sigma_{\phi \phi} 
= {g^2 \over 16 \pi s} \simeq {g^2 \over 64 \pi m_\phi^2}
\quad.
\label{eq:cross1} 
\end{equation}

A limit on $m_\phi$ and $g$ can be obtained by demanding that the mean free
path of the particle $\phi$, $\lambda_\phi$,  is in the interval 
$1~\mbox{kpc} < \lambda_{\phi} < 1~$Mpc \cite{Spergel}.  
Hence, we have:

\begin{equation}
\lambda_{\phi} = {1 \over \sigma_{\phi \phi} n_\phi} 
= {m_\phi \over \sigma_{\phi\phi} \rho^{h}_\phi} 
\quad,
\label{eq:cross11} 
\end{equation}
where $n_\phi$ and $\rho^{h}_\phi$ are, respectively, 
the number and mass density of  $\phi$ particles in the halo. 
Eqs. (\ref{eq:cross1}) and (\ref{eq:cross11})   imply:

\begin{eqnarray}
\sigma_{\phi \phi} & = & 2.1 \times 10^{3} 
\left({m_\phi \over \gev}\right) 
\left({\lambda_\phi \over \mpc} \right)^{-1}\nonumber\\ 
&\times & \left({\rho^{h}_\phi \over 0.4~\mbox{\gev cm}^{-3}} \right)^{-1}
~~\gev^{-2}
\quad,
\label{eq:cross2}
\end{eqnarray} 
which, in turn, leads to:

\begin{equation}
m_\phi= 13~g^{2/3}
\left({\lambda_{\phi} \over \mpc} \right)^{1/3} 
\left({\rho^{h}_\phi \over 0.4~\mbox{\gev cm}^{-3}}\right)^{1/3}\mev 
\quad.
\label{mass}
\end{equation}

In what follows, we shall see how the requirement that $\Omega_{\phi} h^2 
\simeq 0.3$, i.e. that the phion is a suitable DM candidate, and
that the phion is able to explain small scale
structure, leads to bounds on the couplings $g$ and $g^{\prime}$. 

\section{Phion density estimate}
\label{sec:estimate}

In the ensuing discussion, we shall bear in mind the most recent lower 
bound on the mass of the SM Higgs boson, as it emerges from the combined 
LEP data gathered at energies up to $205.9~ \gev$, 
$m_h > 112.3~ \gev$ at $95\%$ confidence level \cite{junk}. On the other hand, 
SM precision data indicates that $m_h < 188~ \gev$ at $95\%$ 
confidence level \cite{junk} and $m_h \le 306~ \gev$ at $99\%$ 
confidence level \cite{PDG}.

If $g^{\prime}$ is sufficiently small, phions decouple early  in the 
thermal history of the Universe and are diluted by subsequent entropy 
production.
In Ref. \cite{Bento1}, we have considered out-of-equilibrium phion production 
via inflaton 
decay in the context of $N=1$ Supergravity inflationary models (see. 
e.g. \cite{Bento2} and references therein).

On the other hand, for certain values of the coupling $g^\prime$, it is possible
that $\phi$ particles are in thermal equilibrium with ordinary matter. 
In order to determine whether this is the case, we will make the usual 
comparison between the thermalization rate $\Gamma_{th}$ and the expansion 
rate of the Universe $H$.

The thermalization rate is given by
\begin{equation}
\Gamma_{th} = n <\sigma_{ann} v_{rel}>
\quad,
\end{equation}
where 
$n = 1.2 \times T^3/\pi^2$  is 
the density of relativistic phions and $< \sigma_{ann} v_{rel}> $ is the 
annihilation
cross section averaged over relative velocities. 
On the other hand,  the expansion rate is given by:
\begin{equation}
H = \left({4 \pi^3 g_\ast \over 45}\right)^{1/2} {T^2 \over M_{P}} = 
1.66 \times g_\ast^{1/2} {T^2 \over M_{P}}
\quad.
\label{hubble}
\end{equation}

At temperatures above the electroweak phase transition, 
$T_{EW} \simeq 300~\gev$, a typical value in many extensions of the SM where 
one hopes to find the required features to achieve successful baryogenesis 
\cite{Kajantie,Laine}, the
order parameter (the vacuum expectation value of the Higgs field) vanishes, 
$v(T) = 0$, and therefore the $\phi\phi h$ coupling is non-operative. 
However, as mentioned above, this interaction term has its origin in the 
4-point coupling, $\phi \phi h h$, which can bring, at high temperatures, the
phion-Higgs system into thermal equilibrium.
Using the temperature as the center-of-mass energy, the cross section is given
by:
\begin{equation}
\sigma_{ann} v_{rel} \simeq {g^{\prime 2} \over 32 \pi T^2}
\quad,
\end{equation} 
which implies that phions are in thermal equilibrium for temperatures smaller
than 
\begin{equation}
T_{eq} \simeq {g^{\prime 2} M_P \over 32 \pi^3 g_\ast^{1/2}}
\quad. 
\end{equation}
Therefore,  phions will never be in thermal equilibrium
before the
electroweak phase transition if $g^{\prime} \lsim 10^{-7}$. 

Thermal equilibrium can be achieved just below $T_{EW}$, when 
the trilinear coupling is operative, if $g^\prime$ is such that
the thermalization rate, $\Gamma_{th}$, is larger than the 
Hubble expansion rate. We now compute 
the required bounds on $g^\prime$.

The phion annihilation cross section at high energies ($T \gsim m_h$) 
is given by a relativistic Breit-Wigner resonance formula:
\begin{equation}
\sigma_{ann} v_{rel} = {
4 \pi (s/m_h^2) \Gamma(h \rightarrow \phi \phi) \Gamma_h
\over
(s - m_h^2)^2 +
m_h^2\Gamma_h^2}
\quad,
\label{sigma}
\end{equation}
where $\Gamma_h$ is the total Higgs decay rate. At the resonance peak 
($s = m_h^2$) it becomes
\begin{equation}
\sigma_{ann} v_{rel} = {4 \pi \over m_h^2} BR(h \rightarrow \phi \phi)
\quad.
\end{equation}

Using
\begin{equation}
\Gamma(h\rightarrow \phi\phi)={g{^\prime}^{2} v^{2} 
(m_h^2 - 4  m_{\phi}^2)^{1/2} \over 32 \pi m_h^2}
\quad,
\label{eq:gphi}
\end{equation}
we obtain the decoupling temperature in the limit $m_h \gg m_\phi$

\begin{equation}
T_D \simeq 150 {\Gamma_h m_h^3 \over g^{\prime 2} M_P v^2} 
\quad.
\end{equation}
This implies that, in order to have 
a decoupling temperature of the order of the
Higgs mass, the coupling constant should be fairly small:
\begin{equation}
g^{\prime} \simeq 10^{-10}
\quad,
\end{equation}
where we have introduced the SM value 
of $\Gamma_h = 3.2 ~\mev$, obtained from
the code HDECAY \cite{hdecay}, used to compute the width of
a $115 ~\gev$ Higgs boson.

Therefore, if $g^{\prime} \geq 10^{-10}$, the phions will be brought 
into thermal equilibrium right after the electroweak phase transition.
If this is the case, there are two possible scenarios depending whether
they decouple while relativistic or otherwise.

\begin{figure}
\centerline{\epsfysize=5cm \epsfbox{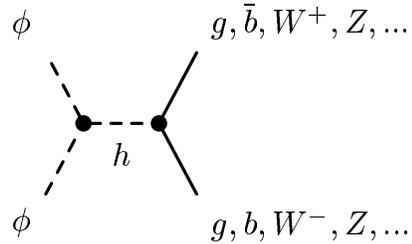}}
\caption{Feynman diagramm for phion annihilation via Higgs exchange.}
\label{fig:phiong}
\end{figure}

In order to study these scenarios, we need to determine the
decoupling temperature at $T \simeq m_\phi \ll T_{EW}$, in which case the 
phion annihilation cross section involves virtual Higgs exchange 
($h^{\ast}$), as in  Figure~\ref{fig:phiong}, and is given 
by \cite{Burgess}:

\begin{equation}
\sigma_{ann} v_{rel} ={8 {g^{\prime}}^2 v^2 \over (4 m_\phi^2-m_h^2)^2 +
m_h^2\Gamma_h^2} F_X
\quad,
\label{sigma}
\end{equation}
where

\begin{equation}
F_X=\lim_{m_{h^\ast} \rightarrow 2 m_\phi}
\left({\Gamma_{h^{\ast} X} \over m_{h^\ast}}\right) 
\quad,
\label{fx}
\end{equation}
and $\Gamma_{h^\ast X}$
denotes the width for the decay $h^\ast \rightarrow X$ ($ X \neq \phi \phi$,
since we are dealing with inelastic scattering only), for
$m_{h^\ast} = 2 m_{\phi}$. For the mass range of interest to us, 
$m_\phi \sim 10-100~\mev$, one has $F_X \sim 10^{-13}$ \cite{Gunion}.

In this case, the relationship between the coupling 
constant $g^{\prime}$ and the decoupling temperature $T_D$ is given by 

\begin{equation}
g^{\prime 2} = 5.5 {(m_h/100 \gev)^4 \over (T_D/\mev)}
\quad.
\end{equation}

If $g^{\prime} \leq 0.1$, the phions decouple while relativistic and are 
as abundant as photons. 
Since we are interested in stable light phions, 
it is a major concern avoiding phion 
overproduction if it decouples while relativistic.
In fact,  in this case, there is  an analogue of Lee-Weinberg limit for
neutrinos (see e.g. \cite{Kolb}):

\begin{equation}
\Omega_\phi h^2 \simeq 0.08 {m_\phi \over 1 ~ \ev}
\quad,
\end{equation}
which results in a very stringent bound, $m_\phi \lsim 4 ~\ev$, and implies
that the phion self-coupling constant should satisfy 
$g \lsim 2.5 \times 10^{-10}$, in order to solve 
the small scale structure problem that exists with 
conventional collisionless cold dark matter.

In order that the phions decouple non-relativistically and their
abundance reduces to acceptable levels without fine-tuning the
self-coupling constant, $g^{\prime} > 0.1$ is required. In this case, 
using standard methods to compute 
the phion relic abundance \cite{Burgess,Kolb}, one obtains:

\begin{equation}
\Omega_{\phi} h^2 = {1.07 \times 10^9 x_F \over g_{\ast}^{1/2} M_P 
\vev{\sigma_{ann} v_{rel}}}
\quad,
\label{omega}
\end{equation}
where $g_{\ast}$ denotes the number of degrees of freedom in equilibrium at
annihilation and $x_F \equiv m_{\phi}/T_F$ is the inverse of the freeze-out 
temperature in units of the phion mass. 
The relevant crosss section is the phion annihilation 
cross section involving virtual Higgs exchange, Eq. (\ref{sigma}), with 
$F_X \sim 10^{-13}$ \cite{Gunion}. The freeze-out temperature is determined by 
the solution of the equation

\begin{equation}
x_F \simeq \ln[0.038 (g_{\ast} x_F)^{-1/2} M_P m_{\phi}
\vev{\sigma_{ann} v_{rel}}]
\quad.
\label{xf}
\end{equation}

In order to get $x_F \geq 1$ and apply Eq. (\ref{omega}), one must have 
$g^{\prime} \gsim 1.2$, for $m_\phi = 50 ~\mev$ and $m_h = 115 ~ \gev$. 
However, due
to the smallness of the cross section, the relic abundance of the phion is
several orders of magnitude larger than the observed value. Hence, one
concludes that a phion in this mass range is excluded.

\begin{figure}
\centerline{\epsfysize=7cm \epsfbox{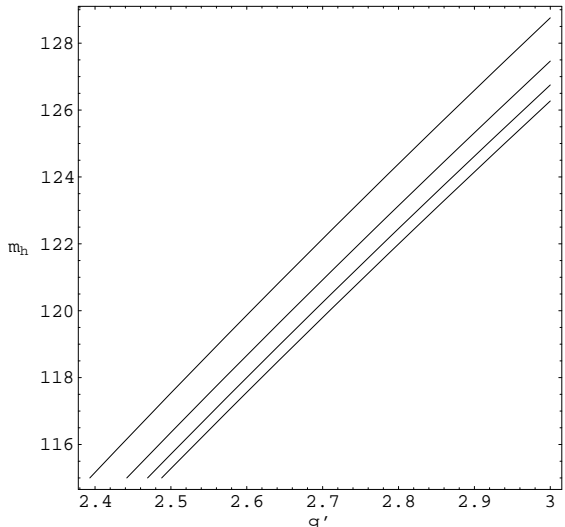}}
\caption{Contour of $\Omega_\phi h^2=0.3$ as a function of $m_h$ (in GeV) and  
$g^\prime$, for $m_{\phi}=0.5~\gev$ (top),~$1.0,~1.5$ and $2~\gev$ (bottom).}
\label{fig:mphiall}
\end{figure}

The smallness of the phion annihilation cross section for $m_\phi \simeq 50
~\mev$ is due to the small factor $F_X \simeq 10^{-13}$. However, this factor
increases significantly with larger phion masses. Considering 
$m_\phi \simeq 1 ~\gev$, then it follows that $F_X \simeq 10^{-7}$.
We find that the requirement $\Omega_{\phi} h^2 \simeq 0.3$ implies that 
$m_{\phi} \gsim 500~\mev$ and $g^{\prime} \gsim 2$, a solution which holds 
only for $m_h \lsim 130~\gev$. Heavier phion and Higgs particles tend to make 
$\Omega_{\phi} h^2 > 0.3$. We have verified that our solutions 
respect the condition $x_F > 3$, meaning that the phion behaves as a cold dark
matter candidate. Of course, our solutions are at the edge of validity of 
perturbation theory.  
Our results are depicted in Figure \ref{fig:mphiall}.

For these large values of the coupling constant, the decay width of the Higgs
into phions is given by:
\begin{equation}
\Gamma(h\rightarrow \phi\phi)= 5.23~\left({m_h \over 115~\gev}\right)^{-1} 
g{^\prime}^{2}
~ \gev.
\end{equation}

Hence, in this case the Higgs width is totally dominated by the 
invisible decay mode and this model can be easily tested at future 
colliders (see e.g. \cite{Eboli}).


\section{Discussion and Conclusions}

In this work, we have derived, from fairly general cosmological arguments, 
bounds on $g^{\prime}$, the 
coupling constant of the Higgs boson to stable scalar particles,
which contribute to Higgs decay 
via invisible channels. These particles, the phions, have all the required 
features to be regarded a successful self-interacting dark matter candidate 
and solve the difficulties of the CDM model on small scales. 

We have found that, for $g^{\prime} \lsim 10^{-10}$, the phions never reach 
thermal equilibrium and hence can only be produced by 
out-of-equilibrium
decay of the inflaton field, as discussed in \cite{Bento1}. In this case, the
phion does not contribute to the invisible Higgs boson decay channel.
For $g^{\prime} \gsim 10^{-10}$, we have found that, if 
$g^{\prime} \lsim 0.1$, the phion decouples while still 
relativistic and a limit for its mass, $m_\phi \lsim
4~\ev$ can be derived, which, in turn, implies in a strong bound on 
the phion self-coupling constant, $g \lsim 10^{-9}$, if 
the phion is required to solve the CDM model problems on small
scales. On the other hand, if $g^{\prime} \gsim 1$,
the phion decouples while non-relativistic; however, its abundance is 
not cosmologically acceptable for phion masses in the range of $50 - 100
~\mev$ due to the small annihilation cross section. For masses in the range of
$0.5 - 2 ~\gev$, we have found that abundances of $\Omega_\phi h^2 \simeq 0.3$
require large values of the coupling $g^{\prime} \simeq 2.5$ and 
$m_h \lsim 130~\gev$. In this scenario,
the Higgs width is dominated by the invisible $h \to \phi \phi$ mode and can be
tested at future colliders.

Regarding the origin of  the phion mass, it is possible that it  arises from
the cancellation between a tachyonic mass, 
due to the dynamics of fields in a more encompassing theory
together with  the contribution from the quartic interaction,
$g^{\prime 2}v^{2}$. 
In fact, in case the only relevant fields are the phion and the Higgs boson,
the phion mass is dominated by a term which arises from electroweak symmetry
breaking, namely, $m_\phi \simeq g' v$, and  if  the phion is  to solve the
small structure problem, this would imply 
$g' \simeq 5 \times 10^{-5}$ which, in turn, brings us to 
the overproduction problem. 

Finally, we comment on the recent observation
\cite{Kaplinghat} that CDM problems at small 
scales can be fixed only through s-channel annihilation of DM 
particles, that is $\sigma_{ann} v_{rel} \propto const$ so that 
$\sigma_{ann} v_{rel} = 2.5 \times 10^{-1} (m/\gev)~\gev^{-2}$. We have 
verified that these conditions cannot be met in 
our model. On more general grounds, it is difficult 
to see how substantial amounts of 
annihilating DM could be obtained from any processes in the early Universe, 
given that the required cross section is so large, although it is clearly a 
quite interesting challenge to build concrete particle physics 
models that exhibit such features.

\vskip 0.3cm

\acknowledgments

\noindent
One of us (O.B.) would like to thank L. Bettencourt, A. Krasnitz, L. 
Teodoro and M.S. Turner for discussions on the subject of this paper. 
R.R. would like to thank A. Natale and T. ter Veldhuis for discussions. 
R.R. is supported by CNPq (Brazil) and FAPESP (S\~ao Paulo). M.C.B. and O.B.
acknowledge the partial support of FCT (Portugal) under the
grant POCTI/1999/FIS/36285.


\end{document}